\documentclass[preprint,proceedings]{rmaa}

\SetYear{2004}
\SetConfTitle{II INTERNATIONAL WORKSHOP ON SCIENCE WITH THE GTC: \\
       1st-light Instruments and the LMT}

\title{Predicted chemical evolution for spiral disks from their
observed rotation curves}

\author{
  M. Moll\'{a},\altaffilmark{1} and
  I. M\'{a}rquez\altaffilmark{2}}

\altaffiltext{1}{C.I.E.M.A.T., Madrid, Spain.}
\altaffiltext{2}{Instituto de Astrof\'{\i}sica de Andaluc\'{\i}a, Granada, Spain.}

\shortauthor{Moll\'{a} \& M\'{a}rquez}
\shorttitle{Predicted chemical evolution for spiral disks}

\suppressfulladdresses

\listofauthors{M. Moll\'{a} \& I. M\'{a}rquez}
\indexauthor{Moll\'{a}, M.}
\indexauthor{M\'{a}rquez, I.}

\abstract{The rotation curves for a sample of 67 spiral galaxies
observed by M\'{a}rquez et al.(2002) have been used as input for the
multiphase chemical evolution model.  By using N[II]/Halfa as
estimator of the oxygen abundance, we constraint the possible models
for each galaxy. We may, then, predict the time evolution of these
galaxies and the present time radial distribution for gas, stars, and
star formation rate surface densities and elemental abundances.}

\resumen{Las curvas de rotaci\'{o}n observadas por M\'{a}rquez et al.(2002) para
una muestra de 67 galaxias espirales han sido usadas como entrada del
modelo de evoluci\'{o}n qu\'{\i}mica multifase. Utilizando  N[II]/Halfa como
estimador de la abundancia de ox\'{\i}geno, podemos limitar los posibles modelos
para cada galaxia. Podemos, por tanto, predecir la evoluci\'{o}n temporal de
estas galaxias y las distribuciones radiales de densidad superficial de  gas, 
estrellas, y tasa de formaci\'{o}n estelar y abundancias elementales para el
momento actual.}

\addkeyword{Galaxies:abundances}
\addkeyword{Galaxies:spirals}
\addkeyword{Galaxies: Star Formation}
\addkeyword{H~II regions}

\begin{document}
\maketitle

\section{Introduction}
\label{sec:intro}

Chemical evolution models are usually applied to our Galaxy, for which
a large number of observations is available. These models may also be
applied to other spiral galaxies (Moll\'{a} et al. 1996; 1999) for
which observational data there exist. They are very useful to
determine the star formation history of the modeled galaxies.

We have applied the multiphase chemical evolution model (Ferrini et
al. 1992) to a sample of 67 galaxies for which the rotation curve, a
necessary input for computing the model, and the estimator
N[II]/H$\alpha$, (which will be used as observational constraint) have
been measured in M\'{a}rquez et al. (2002).  In this paper we show
results for 6 of these galaxies, but see Moll\'{a} \& M\'{a}rquez (in
preparation) for details about models for the whole sample.

\section{Computed Models}
\label{sec:models}

We use the cited measured rotation curves to calculate the total mass
radial distribution in each galaxy with the classical equation:
$M(R)=2.32 10^{5} R. V(R)^{2}$. These distributions, inputs of the
multiphase chemical evolution models, are shown in Fig.~\ref{Fig1},
first row, for 6 selected galaxies. This mass is assumed to be
gas at the initial time which will infall into the equatorial plane
and forms out the disk. The collapse time scale of this gas infall is
defined by the total mass existing within each region.
  
Then, molecular clouds form from the diffuse gas and stars form from
the cloud-cloud collisions. Both formation process rates are
determined by the corresponding efficiencies or probability factors
which depends on environmental effects . As they are not well known
these efficiencies are free parameters and variable for each galaxy.
Once computed the models, we obtain the radial distributions of
abundances which must be compared with observational data in order to
select the best one. The final values of efficiencies will be those
for which the abundances are well reproduced.

The oxygen abundances are estimated by using the empirical relations
given as a function of the parameter $N2=\log{(N[II]/H\alpha)}$ by Van
Zee et al. (1998,VZEE), Raimann et al.(2000,RAI) and Denicol\'{o} et
al.(2002,DEN), respectively:

\begin{eqnarray}
\nonumber
12+ log (O/H)& = & 9.36+1.02 N2  \\
\nonumber
12+ log (O/H)& =& 8.89 (\pm0.07) +0.53(\pm 0.06) N2  \\
\nonumber
12+ log (O/H)& =& 9.12 (\pm 0.05) +0.73 (\pm 0.10) N2  
\label{calibration}
\end{eqnarray}

\section{Results}
\label{sec:results}

\begin{figure*}[!t]\centering
  \vspace{0pt}
  \includegraphics[angle=-90,width=1.8\columnwidth]{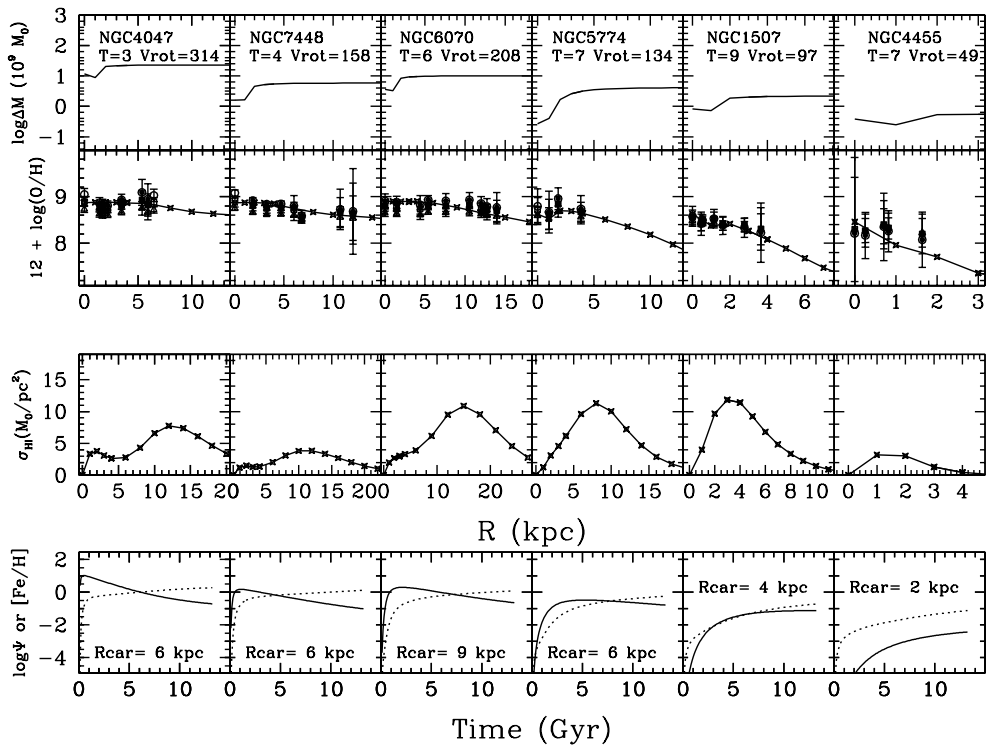}
  \caption{Multiphase chemical evolution model results for 6 galaxies
named in top panels: a) The total mass computed from the rotation
curve, in the first row; b) The oxygen abundance $12 + log(O/H)$ in
the second row; Data -- VZEE (dots), RAI (triangles) and DEN (squares)
-- are shown as filled symbols. c) The predicted surface density of
diffuse gas, $\sigma_{HI} (M_{\odot} pc^{-2})$, in the third row; and
d) The predicted star formation history and age-metallicity
characteristic of each galaxy in the bottom row.}
  \label{Fig1}
\end{figure*}

The radial distributions of oxygen abundances which best reproduce the
data are shown in Fig.~\ref{Fig1}, second row, by a solid line which
is joining the modeled radial regions, represented by $\times$, for
our 6 galaxies.  The data are the filled symbols. \footnote{Open dots
of the same symbols are the values for which the uncertainty in
H$\alpha$ is large, which is considered when H$\alpha < 10$).}

Model results are good in most of galaxies, the radial trend being in
agreement with the observations. It is clear that radial gradients are
systematically steeper for the less massive galaxies and that inner
regions usually show a flatter radial gradient than the observed for
the whole disk.

Two correlations appear among the characteristic oxygen abundance of a
galaxy and its radial gradient and the mass of the galaxy and/or the
morphological type T.  However, the correlations with the rotation
velocity are stronger while the morphological type dependence appears
as secondary.

Once obtained the model able to reproduce the data for each galaxy, we
may consider the time evolution and the present time results given by
this model as representative of each galaxy. 

Thus, in Fig.~\ref{Fig1},
third row, we represent the diffuse gas surface density radial
distributions for the same galaxies, which we consider as model
predictions. They show a maximum along the galactocentric
radius. Larger the rotation velocity, larger the radius where it
appears.

In the last row of Fig.~\ref{Fig1}, the predicted star formation
history (solid line) and the age-metallicity relation (dotted line)
are shown for a radial region given in each panel, characteristic of
each galaxy and equivalent to the Solar Vicinity.  The star formation
is biased toward early times for the more massive galaxies, while it
occurs later for the lowest mass and latest type ones

\section{Conclusions}
\label{sec:final}

The chemical evolution models show to be a very useful tool to
estimate, in a one to one basis, the actual evolutionary track of a
spiral disk, allowing to estimate the star formation history and the
age-metallicity relation.

The multiphase model may be easily applied to a large number of
galaxies. This allows to perform statistical studies and the relation
among the chemical features and on other galactic characteristics as
the Hubble type or the total mass.

The most important conclusion is that the characteristic abundance of
a spiral disk as its radial gradient of oxygen results to be clearly
dependent on rotation velocity. This dependence is stronger than the
corresponding one with the morphological type. This last correlation
also there exists (although shows a larger dispersion), probably as a
consequence of other already known trend: the morphological type of
spirals correlates with the rotation velocity having the less massive
ones the latest morphological types.

The total mass, through its rotation velocity, appears, therefore, as
the very conductor of the evolution in the spiral disks.

\end{document}